\documentclass[10pt]{article}
\usepackage{amssymb}
\usepackage{axodraw}
\usepackage{rotate}
\usepackage{epsfig}
\usepackage{wrapfig}
\usepackage{floatflt}
\begin{document}
%--------------------
\newcommand{\eqnzero}{\setcounter{equation}{0}} 
\newcommand{\mpar}[1]{{\marginpar{\hbadness10000%
                      \sloppy\hfuzz10pt\boldmath\bf#1}}%
                      \typeout{marginpar: #1}\ignorespaces}
\def\mnew{\mpar{\hfil NEW \hfil}\ignorespaces}
\newcommand{\bq} {\begin{equation}}
\newcommand{\eq} {\end{equation}}
\newcommand{\bqa}{\begin{eqnarray}}
\newcommand{\eqa}{\end{eqnarray}}
\newcommand{\nll}{\nonumber\\}

\newcommand{\ip }[1]{u\left({#1}        \right)}    % incoming particle
\newcommand{\iap}[1]{{\bar{v}}\left({#1}\right)}    %    "     anti-p
\newcommand{\op }[1]{{\bar{u}}\left({#1}\right)}    % outgoing p
\newcommand{\oap}[1]{v\left({#1}\right)}            %    "     anti-p

\newcommand{\Litwo}{\mbox{${\rm{Li}}_{2}$}}
\newcommand{\alem}{\alpha_{em}}
\newcommand{\alsS}{\alpha^2_{_S}}
\newcommand{\ds }{\displaystyle}
\newcommand{\sss}[1]{\scriptscriptstyle{#1}}
\newcommand{\sla}[1]{/\!\!\!#1}
%----------------  Masses ...
\def\mgn{mgn}
\def\mw {M_{\sss{W}}}
\def\mws{M_{\sss{W}}^2}
\def\mz {M_{\sss{Z}}}
\def\mh {M_{\sss{H}}}
\def\mhs{M_{\sss{H}}^2}
\def\men{m_{\nu_e}}
\def\mel{m_e}
\def\mup{m_u}
\def\mdn{m_d}
\def\mmn{m_{\nu}}
\def\mmo{m_{\mu}}
\def\mch{mch}
\def\mst{mst}
\def\mtn{mtn}
\def\mta{mta}
\def\mtp{m_t}
\def\mbt{m_b}
\def\mf{m_f}
\def\mv{M_{\sss{V}}}
\def\srt{\sqrt{2}}
\def\qel{Q_f}
\def\qmo{Q_f}
\newcommand{\sqrtL}[3]{\sqrt{\lambda\big(#1,#2,#3\big)}}
%---------------------------    Coupling ...
\newcommand{\vpa}[2]{\sigma_{#1}^{#2}}
\newcommand{\vma}[2]{\delta_{#1}^{#2}}
\newcommand{\af}{I^3_f}
\newcommand{\sqs}{\sqrt{s}}
%----------------------------    stw,ctw ...
\newcommand{\stw}{s_{\sss{W}}  }
\newcommand{\ctw}{c_{\sss{W}}  }
\newcommand{\stws}{s^2_{\sss{W}}}
\newcommand{\stwf}{s^4_{\sss{W}}}
\newcommand{\ctws}{c^2_{\sss{W}}}
\newcommand{\ctwf}{c^4_{\sss{W}}}
%-----------------------------   A-B-C-D functions   ------               
\newcommand{\bff}[4]{B_{#1}\big( #2;#3,#4\big)}             
\newcommand{\fbff}[4]{B^{F}_{#1}\big(#2;#3,#4\big)}        
\newcommand{\scff}[1]{C_{#1}}             
\newcommand{\sdff}[1]{D_{#1}}                 
\newcommand{\dffp}[6]{D_{0} \big( #1,#2,#3,#4,#5,#6;}       
\newcommand{\dffm}[4]{#1,#2,#3,#4 \big) }       
%---                  
\newcommand{\tHmus}{\mu^2}
\newcommand{\epsh}{\hat\varepsilon}
\newcommand{\epsb}{\bar\varepsilon}
%-----------------------------------   Equations, Figures, Tables, Sections...

\newcommand{\chapt}[1]{Chapter~\ref{#1}}
\newcommand{\chaptsc}[2]{Chapter~\ref{#1} and \ref{#2}}
\newcommand{\eqn}[1]{Eq.~(\ref{#1})}
\newcommand{\eqns}[2]{Eqs.~(\ref{#1})--(\ref{#2})}
\newcommand{\eqnss}[1]{Eqs.~(\ref{#1})}
\newcommand{\eqnsc}[2]{Eqs.~(\ref{#1}) and (\ref{#2})}
\newcommand{\eqnst}[3]{Eqs.~(\ref{#1}), (\ref{#2}) and (\ref{#3})}
\newcommand{\eqnsf}[4]{Eqs.~(\ref{#1}), 
          (\ref{#2}), (\ref{#3}) and (\ref{#4})}
\newcommand{\eqnsv}[5]{Eqs.(\ref{#1}), 
          (\ref{#2}), (\ref{#3}), (\ref{#4}) and (\ref{#5})}
\newcommand{\tbn}[1]{Table~\ref{#1}}
\newcommand{\tabn}[1]{Tab.~\ref{#1}}
\newcommand{\tbns}[2]{Tabs.~\ref{#1}--\ref{#2}}
\newcommand{\tabns}[2]{Tabs.~\ref{#1}--\ref{#2}}
\newcommand{\tbnsc}[2]{Tabs.~\ref{#1} and \ref{#2}}
\newcommand{\fig}[1]{Fig.~\ref{#1}}
\newcommand{\figs}[2]{Figs.~\ref{#1}--\ref{#2}}
\newcommand{\figsc}[2]{Figs.~\ref{#1} and \ref{#2}}
\newcommand{\sect}[1]{Section~\ref{#1}}
\newcommand{\sects}[2]{Sections~\ref{#1} and \ref{#2}}
\newcommand{\subsect}[1]{Subsection~\ref{#1}}
\newcommand{\appendx}[1]{Appendix~\ref{#1}}

\def\Cmi{c_{-}}
\def\Cpl{c_{+}}
\def\spr{s'}
\def\betap{\beta_{+}}
\def\betam{\beta_{-}}
\def\betapl{\beta^c_{+}}
\def\betami{\beta^c_{-}}
\def\klmi{k^{-}_1}
\def\klpl{k^{+}_1}
\def\betaf{\beta_f}
\def\ph{\phantom{-}}
\def\phph{\phantom{phantomphantom}}
%--------------------------------------------------  end of file: 
%--------------------------------------------------  cpc_add_def.tex
\newcommand {\uu}[1]{\underline{\underline{#1}}}
\newcommand {\dg}[1]{\bf\color{darkgreen}{#1}}
\newcommand {\db}[1]{\bf\color{darkblue}{#1}}
\newcommand {\dbr}[1]{\bf\color{darkbrown}{#1}}
\newcommand {\rbf}{\red \bf}
\def\mz {M_{\sss{Z}}}
\def\mw {M_{\sss{W}}}
\def\mh {M_{\sss{H}}}
\def\mup{m_u}
\def\mdn{m_d}
\def\srt{\sqrt{2}}
\def\order#1{{\mathcal O}\left(#1\right)}
\def\dd{{\mathrm d}}
\def\MSbar{$\overline{\mathrm{MS}}\ $}
\def\GF {G_{\sss F}}
\def\gw {\Gamma_{\sss W}}
\def\bbl{\bf\blue}
\def\itf{I^{(3)}_f}
\def\thmn{\vartheta_{u\nu}}
\def\thmo{\vartheta_{d l}}
\def\thle{\vartheta_{l}}
\def\mml{m_l}
\def\mmf{m_f}
\def\Ts{T^2}
\def\Us{U^2}
\def\Qs{Q^2}
\def\lk{\hspace*{-3mm}}
\def\sh{s_h}
\def\sm{s_m}
\def\rdmhlms{r_{d_1}}
\def\rdmhllms{r_{d_2}}
\def\rdmhlllms{r_{d_3}}
\def\Nmuu{N_{-}}
\def\Npuu{N_{+}}
%------------------------
\setcounter{page}{0}
\thispagestyle{empty}

$\,$
\vspace*{-1cm}

\begin{flushright}
{{\tt hep-ph/0702115} \\
      February 2007 }
\end{flushright}
\vspace*{\fill}
\begin{center}

{\LARGE\bf  
The three channels of the process $f_1\bar{f}_1 HA\to 0$ in the SANC framework \\[2mm]}
\vspace*{1.5cm}
{\bf D.~Bardin$^{*}$, S.~Bondarenko$^{**}$, L.~Kalinovskaya$^{*}$, G.~Nanava$^{***}$, 
L.~Rumyantsev$^{*}$.}
\vspace*{12mm}

{\normalsize{\it 
$^{*}$ Dzhelepov Laboratory for Nuclear Problems, JINR,        \\
        ul. Joliot-Curie 6, RU-141980 Dubna, Russia;           \\
$^{**}$ Bogoliubov Laboratory of  Theoretical Physics, JINR,   \\ 
        ul. Joliot-Curie 6, RU-141980 Dubna, Russia;           \\
$^{***}$ IFJ im. Henryka Niewodnicza{\'n}skiego, PAN           \\
        ul. Radzikowskiego 152, 31-342 Krak{\'o}w,             \\
on leave from Dzhelepov Laboratory for Nuclear Problems, JINR.
}} 
\vspace*{13mm}

\end{center}

\begin{abstract}
\noindent
In this paper we describe the implementation of the processes $f_1\bar{f}_1 HA\to 0$ 
into the framework of {\tt SANC} system.
Here $A$ stands for a photon and $f_1$ --- for a massless fermion whose mass is neglected everywhere
besides arguments of logarithmic functions. The symbol $\to 0$ means that all 4-momenta flow inwards.
The derived one-loop scalar form factors can be used for any cross channel after an appropriate 
permutation of their arguments $s,t,u$.
We present the covariant and helicity amplitudes for all three possible cross channels. 
For checking of the correctness of the results first of all we observe the independence 
on the gauge parameters and the validity of Ward identity (external photon transversality),
and, secondly, we make an extensive comparison with the other independent calculations. 
\end{abstract}

\vspace*{5mm}

\vfill

\footnoterule
\noindent
{\footnotesize \noindent
Supported by INTAS grant $N^{o}$ 03-51-4007.
}

\newpage

\section{Introduction}
%---------------------
In this article we describe some results obtained with
{\tt SANC} ({\it Support of Analytic and Numerical Calculations for experiments at Colliders}) ---
a system for semi-automatic calculations for various processes of elementary particle
interactions at the one-loop precision level. It is a ``server--client'' system.
The ideology of the calculation, precomputation modules, short user guide of the version 
{\tt V.1.00} and its installation are described in Ref.~\cite{Andonov:2004hi}.
{\tt SANC} client may be downloaded from {\tt SANC} servers Ref.~\cite{homepages}.

In a recent paper~\cite{Bardin:2005dp} we presented an extension of {\tt SANC} tree 
in the $ffbb$ sector, comprising the version {\tt V.1.10}.
In this paper we realize its further extension by inclusion of the process 
$ffHA \to 0$ in all possible cross channels as was pointed yet in section 2.7 
of Ref.~\cite{Andonov:2004hi}. For this reason, we do not change the number  
of {\tt SANC} version, it is still {\tt V.1.10}. 

\begin{floatingfigure}{65mm}
\includegraphics[width=6cm,height=11cm]{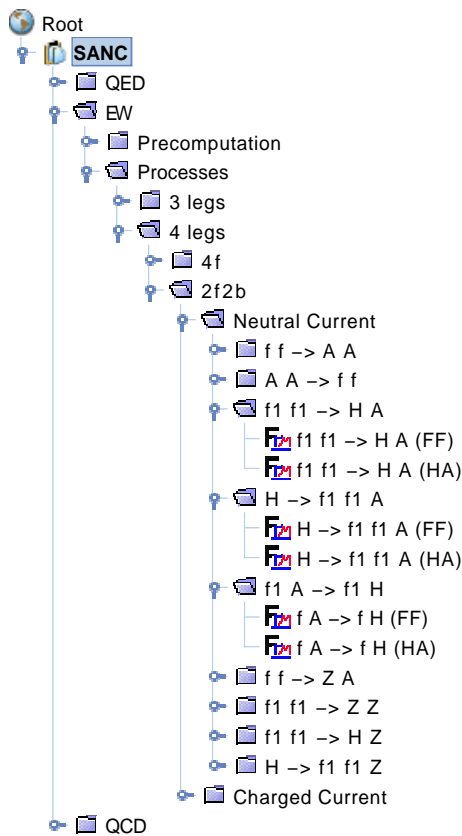}
\caption[New processes in the $ffbb$ sector]
        {New processes in the $ffbb$ sector.}
\label{Processes}
\end{floatingfigure}

The processes $ffHA \to 0$ are interesting for physical applications at LHC (decay channel)
and at ILC, both $e^+e^-$ and $\gamma e$ modes (production channels). There is a rich world 
literature devoted to these processes. We quote below only those papers with which we compare
our numerical results.

The modified branch {\bf 2f2b} for the ``Processes'' tree in the EW part is shown in 
Fig.~\ref{Processes}.
In this paper we consider in detail the process $ffHA\to 0$ in the three channels: \\
$\,$\hspace*{5mm}$\bullet$ annihilation, $f\bar{f}\to H\gamma$; \\
$\,$\hspace*{5mm}$\bullet$ decay, $H\to f\bar{f}\gamma$;\\
$\,$\hspace*{5mm}$\bullet$ and $H$ production at $\gamma e$ colliders, $e\gamma\to e H$.

It contains menus for $f_1\bar{f}_1\to H A$, $H\to f_1\bar{f}_1 A$ and $A f\to H f$ 
which in turn are branched into scalar Form Factors (FF) and Helicity Amplitudes (HA). 
Contrary to a presentation in section 2.7 of Ref.~\cite{Andonov:2004hi}
we extract now the BORN structure containing in the FF $F_{v1}$
and the corresponding expressions for it and HAs for the decay channel $H\to f_1\bar{f}_1 A$ 
are different from Eqs.(55) in parts with $F_{v1}$.

The main objects are FFs. They are the same for all three channels,  
differing only by permutations of arguments $s,t$ and $u$ for different channels. 
For the computation at one-loop, we created symbolic source codes based on \linebreak
FORM3 Ref.~\cite{Vermaseren:2000nd}.
All processes  are implemented at Level 1 of FORM calculations.
% and Level 2, where
%the {\tt s2n.f} package produces the result in the ``Semi Analytic'' mode 
%(see Fig.~20 of Ref.~\cite{Andonov:2004hi}).
%For these processes we also have standalone FORTRAN codes, which are accessible in 
%the {\tt SANC} system. (to discuss with SGB and decide what to write here) 
%We have no Monte Carlo generators for these $2\to 2$ processes.

We pursue three goals: to demonstrate the analytic expression 
for FFs at one-loop level (as an exclusion given their simplicity) and HAs for three channels
of this process (in the spirit of previous {\tt SANC} presentations) and 
to compare results with existing independent calculation.

The paper is organized as follows.
In section~\ref{Amplitudes} we demonstrate an analytic expression for the covariant amplitude (CA) 
at one-loop level in the annihilation channel and give explicit expressions for all FFs. 
Then we give HAs for all three channels available in the {\tt SANC V.1.10}.

In section~\ref{Num_results_comp} we show numerical results (computed by software $s2n$)
and comparison with the other independent calculations: for the decay channel $H\to f_1\bar{f}_1 A$
at the tree level Ref.~\cite{Boos:2004kh} and Ref.~\cite{Yuasa:1999rg} and in the resonance
approximation at the one-loop level Ref.~\cite{Sjostrand:2006za}. For two channels
$e^{+}e^{-} \to H A $ and $e A \to e H $ we compare with one-loop level calculations 
of Ref.~\cite{Barroso:1985et}, Ref.~\cite{Abbasabadi:1995rc} and Ref.~\cite{Djouadi:1996ws} 
(former process) 
and of Ref.~\cite{Gabrielli:1997ix} and Ref.~\cite{Banin:1998ap} (latter one) in the wide ranges of cms 
energies and Higgs masses.
 
\clearpage

\section{Amplitudes \label{Amplitudes}}
%--------------------------------------
We begin with a schematic represantation of the diagram
of the process $\bar{f}(p_1)f(p_2)\gamma(p_3)H(p_4)\to0$
with

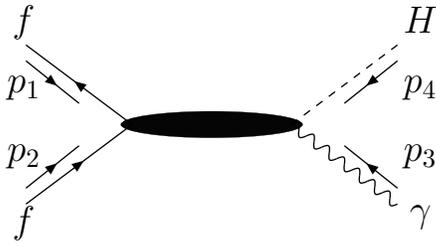
\begin{floatingfigure}{75mm}
\[
\begin{picture}(125,80)(210,10)
   \GOval(270,40)(34,5)(90){0.02}
\ArrowLine(240,40)(200,70)
\ArrowLine(200,10)(240,40)
\DashLine(340,70)(300,40){3}
\Photon(300,40)(340,10){2}{7}

\ArrowLine(200,16)(220,32)
\ArrowLine(200,64)(220,48)
\ArrowLine(340,16)(320,32)
\ArrowLine(340,64)(320,48)

\Text(200,54)[]{\Large $p_1$}
\Text(200,28)[]{\Large $p_2$}
\Text(350,28)[]{\Large $p_3$}
\Text(350,54)[]{\Large $p_4$}

\Text(200,80)[]{\Large$\bar{f}$}
\Text(200,3)[]{\Large $f$}
\Text(350,5)[]{\Large $\gamma$}
\Text(350,80)[]{\Large$H$}
\end{picture}
\]
\vspace*{-5mm}
\caption
[The $\bar{f}f\gamma H\to 0$ process.]
{The $\bar{f}f\gamma H\to 0$ process.\label{DiagrammaffHACA}}
\end{floatingfigure}

%\vspace*{10mm}
\noindent
all 4-momenta incoming $p_1+p_2+p_3+p_4=0$.
We will consider three cross channels of the process:
annihilation, decay and $H$ production.
For all three channels we can write down almost unique CA. 

Below we give it in the form
corresponding to the annihilation channel, $f(p_2)\bar{f}(p_1)\to H(-p_4)\gamma(-p_3)$.
It might be easily converted into any other channel by a proper permutation of external 
4-momenta.

This is not the case, however, for the HAs. The latter has to be recomputed for all three channels.

\subsection{Covariant amplitude of the process $\bar{f}fH\gamma\to 0$}
%---------------------------------------------------------------------
There are eight transversal in photonic 4-momentum structures, 4 vector and 4 axial ones:
\bqa
{\cal A}_{\sss ffH\gamma}&\hspace*{-4mm}=\hspace*{-4mm}&
\iap{p_1}\Bigg\{\Bigg[\frac{(p_2)_{\nu}}{T^2+m^2_f}-\frac{(p_1)_{\nu}}{U^2+m^2_f}
 -\frac{1}{2}\hspace*{-1mm}\left(\frac{1}{T^2+m^2_f}+\frac{1}{U^2+m^2_f}\hspace*{-1mm}\right)
                                    \sla{p_3}\gamma_{\nu}\Bigg] F_{v1}(\Qs,\Ts,\Us)
\nll
 &&\hspace{11mm}+\,i\,\Bigg[(U^2+m^2_f)(p_2)_{\nu}-(T^2+m^2_f)(p_1)_{\nu}\Bigg]\gamma_5 F_{a1}(\Qs,\Ts,\Us)
\\
 &&\hspace{11mm}+\sla{p_3}\gamma_{\nu}\, \Bigg[F_{v2}(\Qs,\Ts,\Us)+\gamma_5 F_{a2}(\Qs,\Ts,\Us)  \Bigg]
\nll
 &&\hspace{11mm}+\,i\,\Bigg[\sla{p_3}(p_1)_{\nu}-\frac{1}{2}(U^2+m^2_f)\gamma_{\nu}  \Bigg] 
                                         \Bigg[F_{v3}(\Qs,\Ts,\Us)+\gamma_5 F_{a3}(\Qs,\Ts,\Us)  \Bigg]
\nll[1mm]
 &&\hspace{11mm}+\,i\,\Bigg[\sla{p_3}(p_2)_{\nu}-\frac{1}{2}(T^2+m^2_f)\gamma_{\nu}\,\Bigg] 
                                         \Bigg[F_{v4}(\Qs,\Ts,\Us)+\gamma_5 F_{a4}(\Qs,\Ts,\Us)  \Bigg]
%\nll
% &&\hspace{11mm} 
\Bigg\}\ip{p_2}\varepsilon^{\gamma}_{\nu}(p_3).
\nonumber
\label{uniCA}
\eqa
%---

All 4-momenta are incoming and the usual Mandelstam invariants and in Pauli metric $p^2=-m^2$ one has:
\bqa
  (p_1+p_2)^2=Q^2=-s, \qquad (p_2+p_3)^2=T^2=-t, \qquad  (p_2+p_4)^2=U^2=-u;
\eqa
Note, that this representation differs slightly from Eq.~(53) of Ref.~\cite{Andonov:2004hi},
as we found appropriate to construct the Born-like FF as given in Eq.~(58).

For the  process under interest the CA at one-loop order has the form:
\bqa
{\cal A}^{\rm{Born~+~1-loop}}={\cal A}^{\rm{Born}}[{\cal{O}}(\mf)]
                             +{\cal A}^{\rm{1-loop}}[{\cal{O}}(\alpha)]
                             +{\cal A}^{\rm{1-loop}}[{\cal{O}}(\mf\alpha)]\,.
\eqa
For this reason Born amplitude typically contribute less than the one-loop one.
Since the $f$ can not be a top quark, for all channels one may neglect the third term.
Then for the squared amplitude one has:
\bqa
|{\cal A}^{\rm{Born~+~1-loop}}|^2  \longrightarrow |{\cal A}^{\rm{Born}}[{\cal{O}}(\mf)]
                             +{\cal A}^{\rm{1-loop}}[{\cal{O}}(\alpha)]|^2.
\eqa 
For first generation fermions even ${\cal A}^{\rm{Born}}$ should be neglected. Note, that for the same
reason the QED one-loop and the bremsstrahlung corrections do not contribute.

\subsection{Diagrams contributing to ${\cal A}^{\rm{1-loop}}[{\cal{O}}(\alpha)]$, form factors}
Here we discuss which one-loop Feynman diagrams contribute to ${\cal A}^{\rm{1-loop}}[{\cal{O}}(\alpha)]$,
not suppressed by Yukawa coupling. For definiteness, we discuss annihilation channel.
There are only a few of them:
\begin{enumerate}
\item ``Right'' three-boson ($B\gamma H,\;B=\gamma,Z$) vertex, see Fig.6 of Ref.~\cite{Bardin:2005dp}.
The diagram with $B=\gamma$ leads to a Coulomb singularity for the decay and production channels.
\item Boxes of T1 and T3 topologies with virtual $W$ boson,  see Fig.15 of Ref.~\cite{Andonov:2004hi}.
\item Box of T5 topology with virtual $Z$ boson,             see Fig.16 of Ref.~\cite{Andonov:2004hi}.
\item Associated $WW$ and $ZZ$ vertices of the topology BFB, see Fig.10 of Ref.~\cite{Andonov:2004hi}.
\end{enumerate}
As was motivated above, we keep the two Born diagrams of the kind shown in Fig.~1 
Ref.~\cite{Bardin:2005dp}.

Every FF is presented as the sum over the gauge index
($1=\xi_{\sss{A}},2=\xi_{\sss{Z}},3=\xi_{\sss{W}}\equiv\xi,4=\mbox{without}\,\xi$):
\bqa
F_{v(a)i}(-s,-t,-u)=\sum_{k=1,4}{F_{v(a)ik}(-s,-t,-u)}.
\eqa
Obviously, $k=1$ does not contribute in massless case. 
FFs for $k=2,3,4$ are rather compact, they are shown in the next section.

\subsection{One-loop form factors}
%---------------------------------
In the limit $\mf \to 0$ FFs with the gauge index $2$ take a simple form: 
\bqa
F_{v32}(-s,-t,-u) &=& \qmo\frac{v_f^2+a_f^2}{2} \, F_2(-s,-t,-u)\,,
\nll 
F_{v42}(-s,-t,-u) &=& \qmo\frac{v_f^2+a_f^2}{2} \, F_2(-s,-u,-t)\,,
\nll[1mm]
F_{a32}(-s,-t,-u) &=& \qmo v_f\, a_f\, F_2(-s,-t,-u)\,,
\nll[3mm]
F_{a42}(-s,-t,-u) &=& \qmo v_f\, a_f\, F_2(-s,-u,-t)\,,
\eqa
where an auxiliary function was introduced: 
\bqa
F_2(\Qs,\Ts,\Us) &=& \frac{\stw}{\ctw^3} \frac{\mz}{\Qs\Us} \biggl\{
           \left(\mz^2+\Us\right) \frac{1}{\Us} \, C_{d_0c_0}(\Ts,\Us,-\mh^2;\mz)
\nll
 &&   +\left(\mz^2+\Us\right) \left(\mh^2+\Us\right)\frac{1}{\Us}  \scff{0}(-\mh^2,-\mf^2,\Us;\mz,\mz,\mf)
\nll
&&+\left[\Qs-\Us+\mz^2\left(1+\frac{\Qs}{\Us}+2\frac{\Us}{T^2+\mh^2}\right)\right]
\scff{0}(-\mh^2,-\mf^2,\Ts;\mz,\mz,\mf)
\nll
&&-\frac{2\Qs}{T^2+\mh^2}\Big[\fbff{0}{-\mh^2}{\mz}{\mz}-\fbff{0}{\Ts}{\mz}{\mf}\Big] \biggr\},
\eqa
with a combination of Passarino--Veltman functions
which is explicitly free off fermionic mass singularities:
\bqa
C_{d_0c_0}(\Ts,\Us,-\mh^2;\mz)&\lk=\lk&\big[-\Ts\Us-\big(\Ts+\Us\big)\mz^2\big]
\sdff{0}(0,-\mf^2,-\mh^2,-\mf^2,\Ts,\Us;\mf,\mf,\mz,\mz)
\nll
&&-\Ts\scff{0}(0,-\mf^2,\Ts;\mf,\mf,\mz)
-\Us\scff{0}(0,-\mf^2,\Us;\mf,\mf,\mz)\,.
\eqa

The FFs with gauge index $4$ are:
\bqa
&&F_{v34}(\Qs)=F_{v44}(\Qs)=2\,\sum_{i}\,\stw\,c_i\frac{m_i^2}{\mw}
  \left(4\frac{Q_f Q^2_i\stw^2}{\Qs}+\frac{v_f Q_i v_i}{\ctw^2}\,\frac{1}{\Qs+\mz^2}\right) F_{4i}\,,
\nll
&&F_{a34}(\Qs)=F_{a44}(\Qs)=2\,\sum_{i}\,\stw\,c_i\frac{m_i^2}{\mw}
  \left(                             \frac{a_f Q_i v_i}{\ctw^2}\,\frac{1}{\Qs+\mz^2}\right) F_{4i}\,,
\eqa
%---
with the auxiliary function,
\bqa
F_{4i}(\Qs)&=&\left(\frac{1}{2}-2\frac{m^2_i}{\Qs+\mh^2}\right)\scff{0}(0,-\mh^2,\Qs,m_i,m_i,m_i)
\nll
&&+\frac{1}{\Qs+\mh^2}\left[1-\frac{\Qs}{\Qs+\mh^2}
                   \Big(\fbff{0}{-\mh^2}{m_i}{m_i}-\fbff{0}{\Qs}{m_i}{m_i}\Big)\right],
\eqa

The FFs with gauge index $3$ are more cumbersome:
\bqa
F_{v33}(-s,-t,-u)&=&Q_f F_{q3}(-s)+v_f F_{va3}(-s)+F_{3}(-s,-t,-u)\,,
\nll[2mm]
F_{v43}(-s,-t,-u)&=&Q_f F_{q3}(-s)+v_f F_{va3}(-s)+F_{3}(-s,-u,-t)\,,
\nll[2mm]
F_{a33}(-s,-t,-u)&=&a_f F_{va3}(-s)+F_{3}(-s,-t,-u)\,,
\nll[2mm]
F_{a43}(-s,-t,-u)&=&a_f F_{va3}(-s)+F_{3}(-s,-u,-t)\,,
\eqa
and one needs three auxiliary functions to define them:
\bqa
F_{q3}(\Qs)&=&2\frac{\stw^3}{\mw}\biggl\{-2\frac{\mw^2}{\Qs}\left(4-\frac{\mh^2+6\mw^2}{\Qs+\mh^2}\right)
          \scff{0}(0,-\mh^2,\Qs;\mw,\mw,\mw)
\nll
&&+\frac{\mh^2+6\mw^2}{\Qs+\mh^2}
\bigg[\frac{1}{\Qs+\mh^2}\bigg(\fbff{0}{-\mh^2}{\mw}{\mw}-\fbff{0}{\Qs}{\mw}{\mw}\bigg)-\frac{1}{\Qs}\bigg]\bigg\},
\nll[2mm]
F_{va3}(\Qs)&=&\frac{\stw}{\mw}\bigg\{\frac{1}{\Qs+\mz^2}
\bigg[\bigg(\frac{1}{\ctw^2}-6-\bigg(\frac{1}{\ctw^2}-12\ctw^2\bigg)\frac{\mz^2}{\Qs+\mh^2}\bigg)
\nll[2mm]
&&\times\mw^2\scff{0}(0,-\mh^2,\Qs;\mw,\mw,\mw)  
\nll[2mm]
&&+\bigg(\frac{1}{2\ctw^2}-1+\bigg(\frac{1}{2\ctw^2}-6\ctw^2\bigg)\frac{\mz^2}{\Qs+\mh^2}\bigg)
\nll
&&\times\bigg(\frac{\mz^2}{\Qs+\mh^2}\bigg(\fbff{0}{-\mh^2}{\mw}{\mw}-\fbff{0}{\Qs}{\mw}{\mw}\bigg)+1\bigg)
\bigg]
\nll
&&+\frac{1}{\Qs+\mh^2}\bigg[\bigg(\frac{1}{\ctw^2}-2\bigg)\mw^2\scff{0}(0,-\mh^2,\Qs;\mw,\mw,\mw)
\nll
&&+\frac{\mz^2}{\Qs+\mh^2}\bigg(-\frac{1}{2\ctw^2}+6\ctw^2+\bigg(1-\frac{1}{2\ctw^2}\bigg)\frac{\mh^2}{\mz^2}\bigg)
\nll
&&\times\bigg(\fbff{0}{-\mh^2}{\mw}{\mw}-\fbff{0}{\Qs}{\mw}{\mw}\bigg)-\frac{1}{2\ctw^2}+1
\bigg]
\bigg\},
\nll[2mm]
F_{3}(\Qs,\Ts,\Us)&=&\frac{1}{2}\stw\mw
\bigg\{\frac{1}{\Qs U_s^4}
\bigg[\big(\Us+\mw^2\big)\bigg(\Qs\big(\Us+\mw^2\big)+\mw^2\mh^2\bigg)\,d_{0\rm{aux}}(\Qs,\Us)
\nll
&&
+\bigg(\big[\big(\Us+\mw^2\big)\mw^2+\Qs\Ts\big]\mh^2+2\Qs\Us\mw^2+\Qs\big(\Ts-\mw^2\big)^2\bigg)\,d_{0\rm{aux}}(\Qs,\Ts)
\nll
&&
      +Q_s^4 \scff{0}(-\mmo^2,-\mmo^2,\Qs;\mw,0,\mw)
\nll[1mm]
&&
      +\big(\Qs+\mh^2\big)\big[\Qs+2\big(\Us+\mw^2\big)\big]\scff{0}(0,-\mh^2,\Qs;\mw,\mw,\mw)
\nll
&&
+\bigg(\Qs\big(\Ts+\mh^2\big)+\big(\Us+\mw^2\big)(\Us-\Qs)+2\frac{\mw^2\Qs\Us}{\Ts+\mh^2}\bigg)
\nll
&&
 \times\scff{0}(-\mh^2,-\mmo^2,\Ts;\mw,\mw,0)
\nll[1mm]
&&
      -\big(\Us+\mh^2\big)\big(\Us+\mw^2\big)\scff{0}(-\mh^2,-\mmo^2,\Us;\mw,\mw,0)
\nll[1mm]
&&
      +\Ts\big(\Qs+\Us+\mw^2\big)\scff{0}(0,-\mmo^2,\Ts;\mw,\mw,0)
\nll[1mm]
&&
      +\Us(\Us+\mw^2)\scff{0}(0,-\mmo^2,\Us;\mw,\mw,0) 
\bigg]
\nll
&&
      +\frac{2}{\Us}\,\frac{1}{\Ts+\mh^2}\bigg(\fbff{0}{-\mh^2}{\mw}{\mw}-\fbff{0}{\Ts}{\mw}{0}\bigg) 
\bigg\}.
\eqa
%---
Here two two more auxiliary functions are introduced:
\bqa
d_{0\rm{aux}}(\Qs,\Ts)=d_0(-\mmo^2,-\mmo^2,0,-\mh^2,\Qs,\Ts;\mw,0,\mw,\mw)\,.
\nll[2mm]
d_{0\rm{aux}}(\Qs,\Us)=d_0(-\mmo^2,-\mmo^2,-\mh^2,0,\Qs,\Us;\mw,0,\mw,\mw)\,,
\eqa
The complete analytic results were presented in the literature earlier, see e.g. 
Ref.~\cite{Djouadi:1996ws} and references therein. 
The aim of presentation of this section is to show once typical SANC result for the FF in terms 
of only {\em scalar} Passarino--Veltman functions.

%\clearpage

\subsection{Helicity amplitudes}
%-------------------------------
In this section we present the HAs for all three channels.
\subsubsection{Annihilation channel $\bar{f}f\to\gamma H$}
%---------------------------------------------------------
In this case permutation of 4-momenta with respect to CA~(\ref{uniCA}) is very simple.
\vspace*{2mm}
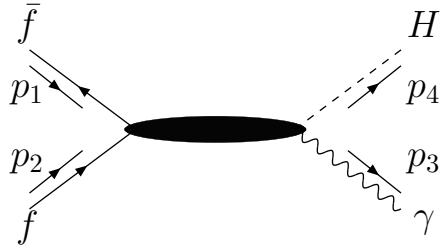
\begin{floatingfigure}{75mm}
\[
\begin{picture}(125,80)(230,-20)
   \GOval(270,40)(34,5)(90){0.02}
\ArrowLine(240,40)(200,70)
\ArrowLine(200,10)(240,40)
\DashLine(340,70)(300,40){3}
\Photon(300,40)(340,10){2}{7}

\ArrowLine(200,16)(220,32)
\ArrowLine(200,64)(220,48)
\ArrowLine(320,32)(340,16)
\ArrowLine(320,48)(340,64)

\Text(200,54)[]{\Large $p_1$}
\Text(200,28)[]{\Large $p_2$}
\Text(350,28)[]{\Large $p_3$}
\Text(350,54)[]{\Large $p_4$}

\Text(200,80)[]{\Large$\bar{f}$}
\Text(200,3)[]{\Large $f$}
\Text(350,5)[]{\Large $\gamma$}
\Text(350,80)[]{\Large$H$}
\end{picture}
\]
\vspace*{-15mm}
\caption
[Schematic representation of one-loop Feynman diagrams for the annihilation channel.]
{Schematic representation of one-loop Feynman  diagrams for the annihilation channel.
\label{DiagrammaffHA}}
\end{floatingfigure}

\vspace*{5mm}

Convert for channel $\bar{f}(p_1)f(p_2)\to\gamma(p3)H(p4)$:

\[
\begin{array}{llll}
& p_1 & \to &~~p_1, \\
& p_2 & \to &~~p_2, \\
& p_3 & \to & -p_3, \\
& p_4 & \to & -p_4. \\
\end{array}
\]

\clearpage

The set of corresponding HAs for the case reads:
\bqa
{\cal H}_{\pm\pm\pm} &=&\mp k_0
              \biggl[  \frac{-Z_4(\mh)\betap/2+s \beta^2}{Z_1(\mf)Z_2(\mf)} F_{v1}(-s,-t,-u) 
\nll &&
\mp s\beta F_{a1}(-s,-t,-u) 
                          - \betap\left( F_{v2}(-s,-t,-u)  \mp F_{a2}(-s,-t,-u) \right)
\nll &&
                      + \mmf\left( F_{v3}(-s,-t,-u) \mp \beta F_{a3}(-s,-t,-u) \right) 
                      + \mmf\left( F_{v4}(-s,-t,-u) \pm \beta F_{a4}(-s,-t,-u) \right) \biggr],
\nll
{\cal H}_{\pm\pm\mp} &=& \pm k_0 
              \biggl[  \frac{-Z_4(\mh)\betam/2+s \beta^2}{Z_1(\mf)Z_2(\mf)} F_{v1}(-s,-t,-u) 
\nll &&
\mp s\beta F_{a1}(-s,-t,-u) 
                          - \betam\left( F_{v2}(-s,-t,-u) \pm F_{a2}(-s,-t,-u) \right)
\nll &&
                      + \mmf\left( F_{v3}(-s,-t,-u) \mp \beta F_{a3}(-s,-t,-u) \right)
                      + \mmf\left( F_{v4}(-s,-t,-u) \pm \beta F_{a4}(-s,-t,-u) \right) \biggr],
\nll
{\cal H}_{\pm\mp\pm} &=&-k_{+}
       \biggl[
 \frac{2\mmf}{s}\frac{Z_4(\mh)}{Z_1(\mmf) Z_2(\mmf)} F_{v1}(-s,-t,-u)
+\frac{4\mmf}{s}\left( F_{v2}(-s,-t,-u)\mp F_{a2}(-s,-t,-u) \right)
\nll &&
                  -  \betap \left( F_{v3}(-s,-t,-u) \pm \beta F_{a3}(-s,-t,-u) \right)
                  -  \betam \left( F_{v4}(-s,-t,-u) \pm \beta F_{a4}(-s,-t,-u) \right) \biggr],
\nll
{\cal H}_{\pm\mp\mp} &=&-k_{-}
       \biggl[
 \frac{2\mmf}{s}\frac{Z_4(\mh)}{Z_1(\mmf) Z_2(\mmf)} F_{v1}(-s,-t,-u)
+\frac{4\mmf}{s}\left( F_{v2}(-s,-t,-u) \pm F_{a2}(-s,-t,-u) \right)
\nll &&
                  -  \betam \left( F_{v3}(-s,-t,-u) \pm \beta F_{a3}(-s,-t,-u)\right)
                  -  \betap \left( F_{v4}(-s,-t,-u) \pm \beta F_{a4}(-s,-t,-u)\right)  \biggr],
\nll 
\eqa
%---
where the coefficients
\bqa
k_0     = \sin\vartheta_{\gamma}\frac{s_h}{2\sqrt{2}}\,,
\qquad
k_{\pm} = c_{\pm} \frac{s_h\sqrt{s}}{4\sqrt{2}}\,,
\eqa
with
\bqa
s_h=s-\mhs\,,\qquad c_{\pm}=1\pm\cos\vartheta_{\gamma}\,,\qquad \beta_{\pm}=1\pm\beta\,,\qquad 
\beta={\sqrt{\lambda(s,\mf^2,\mf^2)}}/{s}\,,
\eqa
and fermionic propagators and $t,u$ invariants are:
\bqa
Z_1(\mf)&=&\frac{1}{2}Z_4(\mh)\left(1+\beta\cos{\vartheta_{\gamma}}\right),
\nll
Z_2(\mf)&=&\frac{1}{2}Z_4(\mh)\left(1-\beta\cos{\vartheta_{\gamma}}\right),
\nll[2mm]
t&=&\mf^2-Z_2(\mf),
\nll[2mm]
u&=&\mf^2-Z_1(\mf).
\eqa
%---
Here $Z_4(\mh)=s_h$ and $\vartheta_{\gamma}$ is the cms angle of the produced photon 
(angle between 4-momenta $\vec{p}_2$ and $\vec{p}_3$).

\newpage
\subsubsection{Decay channel $H\to\gamma f\bar{f}$}
%--------------------------------------------------
The 4-momenta permutations for this channel are chosen as follows.

%\clearpage
\vspace*{7mm}

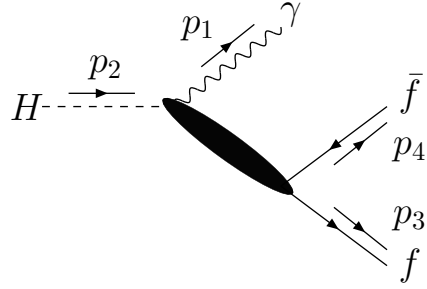
\begin{floatingfigure}{75mm}
\[
\begin{picture}(125,80)(230,10)
\DashLine(210,70)(260,70){3}
\Photon(260,70)(300,100){2}{7}
\Vertex(260,70){1.5}
   \GOval(280,55)(30,5)(54){0.02}
\ArrowLine(340,70)(300,40)
\ArrowLine(300,40)(340,10)
\ArrowLine(320,32)(340,16)
\ArrowLine(320,48)(340,64)
\ArrowLine(270,85)(290,101)
\ArrowLine(220,75)(245,75)
\Text(205,70)[]{\Large $H$}
\Text(305,105)[]{\Large $\gamma$}
\Text(350,10)[]{\Large $f$}
\Text(350,75)[]{\Large $\bar f$}
\Text(235,85)[]{\Large $p_2$}
\Text(270,100)[]{\Large $p_1$}
\Text(350,28)[]{\Large $p_3$}
\Text(350,54)[]{\Large $p_4$}
\end{picture}
\]
\vspace*{-5mm}
\caption
[Schematic representation of one-loop Feynman diagram for the decay channel]
{Schematic representation of one-loop Feynman diagram for the decay channel.}
\label{DiagrammaHAff}
\end{floatingfigure}

Convert for channel $H(p_2)\to\gamma(p_1)f(p_3)\bar{f}(p_4)$: 

\[
\begin{array}{llll}
&\lk p_1 & \to & -p_3,   \\
&\lk p_2 & \to & -p_4,   \\
&\lk p_3 & \to & -p_1,   \\
&\lk p_4 & \to & ~~p_2.
\end{array}
\]

\vspace*{2.5cm}

The HAs for this channel are somewhat similar to the annihilation ones:
\noindent 
\bqa
{\cal H}_{\pm\pm\pm} &=& k_0 \biggl[
          \frac{\betam Z_2(\mh)/2+s \beta^2 }{Z_3(\mf) Z_4(\mf)} F_{v1}(-s,-t,-u)
                                                \pm \beta s  F_{a1}(-s,-t,-u)
\nll &&
                             - \betam \left(F_{v2}(-s,-t,-u)  \mp       F_{a2}(-s,-t,-u)\right)
\nll &&
                               + \mf \left(F_{v3}(-s,-t,-u)\pm \beta F_{a3}(-s,-t,-u)\right)
                               + \mf \left(F_{v4}(-s,-t,-u)\mp \beta F_{a4}(-s,-t,-u)\right) \biggr],
\nll  
{\cal H}_{\pm\mp\mp} &=& k_0 \biggl[
           \frac{\betap Z_2(\mh)/2+s \beta^2}{Z_3(\mf) Z_4(\mf)} F_{v1}(-s,-t,-u)
                                                         \mp \beta s   F_{a1}(-s,-t,-u)
\nll &&
                            - \betap \left(F_{v2}(-s,-t,-u) \mp      F_{a2}(-s,-t,-u)\right)
\nll &&
                               + \mf \left(F_{v3}(-s,-t,-u) \mp\beta F_{a3}(-s,-t,-u)\right)
                               + \mf \left(F_{v4}(-s,-t,-u) \pm\beta F_{a4}(-s,-t,-u)\right) \biggr], 
\nll  
{\cal H}_{\pm\mp\pm} &=& k_{+} \biggl[
                       \mp \frac{2\mf}{s} \frac{Z_2(\mh)}{Z_3(\mf) Z_4(\mf)} F_{v1}(-s,-t,-u)
               \pm   \frac{4\mf}{s}   \left(F_{v2}(-s,-t,-u)  \mp       F_{a2}(-s,-t,-u)\right)
\nll &&               
                          \mp \betap \left(F_{v3}(-s,-t,-u) \pm \beta F_{a3}(-s,-t,-u)\right)
                          \mp \betam \left(F_{v4}(-s,-t,-u) \pm \beta F_{a4}(-s,-t,-u)\right)  \biggr],
\nll   
{\cal H}_{\pm\pm\mp} &=& k_{-} \biggl[
                     \mp  \frac{2\mf}{s}\frac{Z_2(\mh)}{Z_3(\mf) Z_4(\mf)}  F_{v1}(-s,-t,-u)
                 \pm \frac{4\mf}{s}     \left(F_{v2}(-s,-t,-u) \mp       F_{a2}(-s,-t,-u)\right)
\nll &&               
                          \mp \betam \left(F_{v3}(-s,-t,-u) \mp \beta F_{a3}(-s,-t,-u)\right)
                          \mp \betap \left(F_{v4}(-s,-t,-u) \mp \beta F_{a4}(-s,-t,-u)\right) \biggr],
\nll
\eqa
where the coefficients
\bqa
 k_0        &=& \sin\vartheta_{f} \frac{Z_2(\mh)}{2\srt}\,, \nll
 k_{\pm}    &=& c_{\pm}\frac{Z_2(\mh)\sqrt{s}}{4\srt}\,,\nll
\beta_{\pm} &=& 1 \pm \beta\,,\qquad
\beta       = \sqrt{1-\frac{4 \mf^2}{s}}.
\eqa
and the propagators of radiating fermions and $s,t,u$ invariants are: 
\bqa
Z_3(\mf)&=&\frac{1}{2}Z_2(\mh)\left(1+\beta\cos{\vartheta_{f}}\right),
\nll
Z_4(\mf)&=&\frac{1}{2}Z_2(\mh)\left(1-\beta\cos{\vartheta_{f}}\right),
\nll[2mm]
s&=&M^2_{f\bar{f}}\,,
\nll[2mm]
t&=&\mf^2+Z_4(\mf),
\nll[2mm]
u&=&\mf^2+Z_3(\mf).
\eqa
with $Z_2(\mh)=\mhs-s$ and $\vartheta_{f}$ being the fermionic angle $(\vec{p}_3,\vec{p}_1)$
in the rest frame of compound $(\vec{p}_3,\vec{p}_4)$.  

\subsubsection{$H$ production channel $e \gamma \to e H$}
%--------------------------------------------------
For this channel we present HAs for two cases: for limit $\mf\to0$ and exact in $\mf$.

\begin{floatingfigure}{75mm}
\[
\begin{picture}(125,80)(230, 0)
   \GOval(270,40)(34,5)(0){0.02}
\Photon(270,67)(230,97){2}{7}
\DashLine(310,97)(270,67){3}
\ArrowLine(230,-17)(270,13)
\ArrowLine(270,13)(310,-17)
\ArrowLine(240,-20)(260,-4)
\ArrowLine(240,100)(260,84)
\ArrowLine(280,84)(300,100)
\ArrowLine(280,-4)(300,-20)
\Text(230,80)[]{\Large $\gamma$}
\Text(230,5)[]{\Large $ f$}
\Text(310,5)[]{\Large $f$}
\Text(310,80)[]{\Large $H$}
\Text(260,100)[]{\Large $p_1$}
\Text(285,100)[]{\Large $p_4$}
\Text(285,-20)[]{\Large $p_3$}
\Text(260,-20)[]{\Large $p_2$}
\end{picture}
\]
\vspace*{2mm}
\caption
[Schematic representation of one-loop Feynman diagrams for $e \gamma$ channel]
{Schematic representation of one-loop Feynman 
diagrams for $e \gamma$ channel\label{DiagrammaAffH}.}
\vspace*{-2mm}
\end{floatingfigure}

\vspace*{8mm}
Convert for channel $\gamma(p_1)e(p_2)\to e(p_3)H(p_4)$ is: \\
\[
\begin{array}{lllll}
&& p_1 &\to& -p_3,  \\
&& p_2 &\to& ~~p_2, \\
&& p_3 &\to& ~~p_1, \\
&& p_4 &\to& -p_4.
\end{array}
\]
\vspace*{3.5cm}

The small mass limit is remarkably compact:
\bqa 
{\cal H}_{\pm\pm\pm}&=& \pm k_1 \left[ F_{v4}(-s,-t,-u) \mp F_{a4}(-s,-t,-u) \right],
\nll
{\cal H}_{\mp\pm\pm}&=& \mp k_2 \left[ F_{v3}(-s,-t,-u) \mp F_{a3}(-s,-t,-u) \right],
\nll                             
{\cal H}_{\mp\mp\pm}&=&     k_3 \biggl[\frac{\mh^2}{Z_3(m_e)} F_{v1}(-s,-t,-u)
                   \pm k F_{a1}(-s,-t,-u) + 2 s \left(F_{v2}(-s,-t,-u) \pm F_{a2}(-s,-t,-u)\right)\biggr],
\nll
{\cal H}_{\mp\pm\mp}&=&     k_3 \biggl[\left( \frac{s_h}{Z_3(m_e)} - 1 \right) F_{v1}(-s,-t,-u)
                   \pm k F_{a1}(-s,-t,-u) \biggr],
\eqa
with the coefficients
\bqa
k   &=& c_{-} \frac{s s_h}{2},
\qquad\qquad\qquad 
k_1  =  \sin\frac{\vartheta_{f}}{2} s \sqrt{\frac{s_h}{2}}, 
\nll
k_2 &=& c_{+}\sin\frac{\vartheta_{f}}{2}
        \frac{\left(s_h\right)^{{3}/{2}}}{2\sqrt{2}},
\qquad
k_3  =  \cos\frac{\vartheta_{f}}{2} \sqrt{\frac{s_h}{2s}},
\eqa
The exact in $\mf$ fermionic propagators are:
\bqa
Z_2(\mf)&=&s-\mf^2\,,
\nll
Z_3(\mf)&=&\frac{Z_2(\mf)}{2s}\left[s+\mf^2-\mhs+\sqrt{\lambda(s,\mf^2,\mhs)}\cos\vartheta_{f}\right].
\eqa
The Mandelstam variables transform as follows:
\bqa
s& \rightarrow &\frac{1}{2}\left[\left(s-\frac{\mh^2\mf^2}{s}-\mh^2-2\mf^2+\frac{\mf^4}{s}\right)
     -\frac{s-\mf^2}{s}\sqrt{\lambda(s,\mf^2,\mhs)}\cos\vartheta_{f}\right],
\nll[2mm]
t& \rightarrow &s,
\nll[2mm]
u& \rightarrow &\frac{1}{2}\left[\left(s+\frac{\mh^2\mf^2}{s}-\mh^2-2\mf^2-\frac{\mf^2}{s}\right)
     +\frac{s-\mf^2}{s}\sqrt{\lambda(s,\mf^2,\mhs)}\cos\vartheta_{f}\right].
\eqa
The quantity
\bqa
Z_4(\mh)&=&\frac{Z_2(\mf)}{2s}\left[s-\mf^2+\mhs-\sqrt{\lambda(s,\mf^2,\mhs)}\cos\vartheta_{f}\right],
\eqa
represents a would-be Higgs boson propagator, which appears only in numerator, since Higgs does not
radiate photons. For $Z_2(\mf)$ and $Z_4(\mh)$ we use here massless expressions, while for $Z_3(\mf)$ 
--- exact one, since it develops logarithmic mass singularity. For the massive case below, we use all
expressions exact in masses. Also important quantities for this channel are:
\bqa
N_{\pm}=\sqrt{\mmf^2+E_2E_3\mp p_2p_3}\,.
\eqa
In massless case only $N_{+}$ contributes and its limit is
\bqa
N_{+}=\sqrt{\ds\frac{s_h}{2}}\,.
\eqa

The  fully massive case has the following form:
\bqa
\cal{H}_{\pm\pm\pm} &=&  \sin{\frac{\vartheta_{f}}{2}} \biggl[
        \pm \frac{1}{Z_2(\mmf) Z_3(\mmf)}\bigg(
           \sqrt{s}\left[\mh^2-\mmf^2\left( 4 + \rdmhllms \cos{\vartheta_{f}}\right)\right] \Npuu
\nll &&
        +\frac{\mmf}{2} \left( 3 \sh+\mmf^2 (9-\rdmhlms) 
        +\left[\sh+\mmf^2 (3+\rdmhlms) \right] \cos{\vartheta_{f}}\right) \Nmuu\bigg) F_{v1}(-s,-t,-u)
\nll &&
      - \sm\Cpl\left[\frac{\sqrt{s}}{2} (-2+\rdmhlms) \Npuu-\mmf \Nmuu\right]  F_{a1}(-s,-t,-u)
\nll &&
      \pm   2 \left(\sqrt{s} \Npuu-\mmf \Nmuu\right) \left( F_{v2}(-s,-t,-u)\mp F_{a2}(-s,-t,-u) \right)
\nll&&
      \mp    \frac{\mmf}{2} \Big[\sqrt{s} \left(4-\rdmhlms +\rdmhlllms \cos{\vartheta_{f}} \right) \Npuu
                        -\mmf \left(4-\rdmhlms\Cmi\right) \Nmuu \Big] F_{v3}(-s,-t,-u)
\nll &&
      +     \frac{\mmf}{2} \Big[\sqrt{s} \left( \rdmhlms-\rdmhlllms \cos{\vartheta_{f}}\right) \Npuu
                                -\mmf  \rdmhlms \Cmi\Nmuu \Big] F_{a3}(-s,-t,-u)
\nll &&
      \mp                     \left[2 \sqrt{s} \mmf \Npuu-(s+\mmf^2) \Nmuu\right] F_{v4}(-s,-t,-u)
      -                                                     \sm \Nmuu  F_{a4}(-s,-t,-u)
                                                 \biggr],
\\ \nonumber
\cal{H}_{\mp\pm\pm} &=&  \sin{\frac{\vartheta_{f}}{2}}\Cpl\biggl[
  \pm \frac{1}{Z_2(\mmf) Z_3(\mmf)} \left( \frac{\sqrt{s}}{2} \left[ \sh+\mmf^2 (3+\rdmhlms)\right] \Npuu
                                            -s \mmf \rdmhllms \Nmuu\right) F_{v1}(-s,-t,-u)
\nll &&
   +\left(\frac{\sqrt{s}}{2} \left[\sh+\mmf^2 (3-\rdmhlms)\right] \Npuu-\sm\mmf\Nmuu\right) F_{a1}(-s,-t,-u)
\nll &&
      \pm   \frac{1}{2} \left( \sqrt{s} \mmf \rdmhlms \Npuu-s \rdmhlllms \Nmuu \right)
                                  \left(F_{v3}(-s,-t,-u) \mp F_{a3}(-s,-t,-u)\right) \biggr],
\\ \nonumber
\cal{H}_{\mp\mp\pm} &=&  \cos{\frac{\vartheta_{f}}{2}} \biggl[
       \frac{1}{Z_2(\mmf) Z_3(\mmf)} \biggl( \frac{\mmf}{2} \bigl(3 \sh+\mmf^2 (9-\rdmhlms)
\nll &&
                            -\left[ \sh+\mmf^2 (3+\rdmhlms) \right] \cos{\vartheta_{f}}\bigr) \Npuu
           +\sqrt{s} \left[\mh^2-\mmf^2(4 - \rdmhllms \cos{\vartheta_{f}})\right] \Nmuu\biggr) F_{v1}(-s,-t,-u)
\nll &&
     \mp                                  \Cmi \left( \sm\mmf\Npuu
                -\frac{\sqrt{s}}{2} \left[\sh+\mmf^2 (3-\rdmhlms)\right] \Nmuu \right) F_{a1}(-s,-t,-u)
\nll &&
      -                               2 \left( \mmf \Npuu-\sqrt{s} \Nmuu \right) 
                                          \left(F_{v2}(-s,-t,-u) \pm F_{a2}(-s,-t,-u)\right)
\nll &&
      +           \frac{\mmf}{2} \Bigl[\mmf \left(4-\rdmhlms\Cpl\right) \Npuu
                -\sqrt{s} \left( 4-\rdmhlms-\rdmhlllms \cos{\vartheta_{f}}\right) \Nmuu \Bigr] F_{v3}(-s,-t,-u)
\nll &&
      \pm   \frac{\mmf}{2}                \Bigl[ \mmf \rdmhlms \Cpl \Npuu
            -\sqrt{s}  \left( \rdmhlms+\rdmhlllms \cos{\vartheta_{f}}\right) \Nmuu \Bigr] F_{a3}(-s,-t,-u)
\nll &&
      +   \left[ (s+\mmf^2) \Npuu-2 \sqrt{s} \mmf \Nmuu \right] F_{v4}(-s,-t,-u)
      \pm                                                     \sm \Npuu  F_{a4}(-s,-t,-u)
                                         \biggr],             
\\ \nonumber
{\cal H}_{\mp\pm\mp} &=&   \cos{\frac{\vartheta_{f}}{2}} \Cmi \biggl[
        \frac{1}{Z_2(\mmf) Z_3(\mmf)} \biggl(                -s \mmf \rdmhllms  \Npuu
            +\frac{\sqrt{s}}{2} \left[\sh+\mmf^2 (3+\rdmhlms)\right] \Nmuu \biggr) F_{v1}(-s,-t,-u)
\nll &&
 \mp\left(\sm\mmf\Npuu-\frac{\sqrt{s}}{2} \left[\sh+\mmf^2 (3-\rdmhlms)\right] \Nmuu\right) F_{a1}(-s,-t,-u)
\nll &&
 - \frac{1}{2} \left(s \rdmhlllms \Npuu-\sqrt{s} \mmf \rdmhlms \Nmuu \right)
                                  \left(F_{v3}(-s,-t,-u)\mp F_{a3}(-s,-t,-u)\right)
\biggr],
\eqa
where
\bqa
 s_m = s-\mmf^2\,,\quad
 \rdmhlms  = 3-\frac{\mhs- \mmf^2}{s}\,,\quad
 \rdmhllms = 2-\frac{\mhs-2\mmf^2}{s}\,,\quad
 \rdmhlllms= 1-\frac{\mhs-3\mmf^2}{s}\,
\eqa
and $\vartheta_{f}$ is the cms angle of the final fermion.

\section{Numerical results and comparison\label{Num_results_comp}}

 In this section we present results of  numerical calculations and comparisons with
other groups. 
\subsection{ Annihilation channel $f \bar{f} \to \gamma H$ \label{comp_res_ffgammaH}}

 There are many papers devoted to this channel, see for example 
Refs.~\cite{Barroso:1985et}--\cite{Djouadi:1996ws}
and references therein. It is highly non-trivial to realize a tuned comparison of the numerical results,
since the authors do not present the list of input parameters, do not specify calculational scheme, 
although stating an agreement among themselves. Eventually, we found the best to compare with newest
paper Ref.~\cite{Djouadi:1996ws}, namely with Fig.2, showing the $\mh$ dependence of the total cross section
for two values of $\sqrt{s}$=500 (solid line) and 1500 GeV (dashed line).
As can be judged from comparison of their figures with ours, there is a qualitative agreement of the 
cross sections. One should emphasize that we did not find in Ref.~\cite{Djouadi:1996ws} which value of the 
top quark mass on which we observed quite a strong  dependence.

For example, at $\sqrt{s}$=500 GeV and $\mh$=300 GeV, the cross section equals 1.32$\cdot10^{-2}$ fb for
$m_t=174.2$ GeV and 1.89$\cdot10^{-2}$ fb for $m_t=140$ GeV.
\clearpage

$\,$
\vspace*{-1cm}

\begin{figure}[!h]
\begin{center}
\includegraphics[height=12cm,width=7cm,angle=-90]{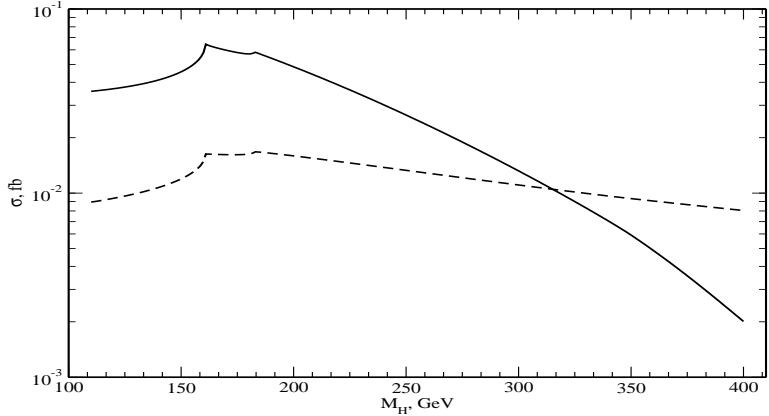}
\caption[Annihilation channel: SANC at the one-loop level]
        {Annihilation channel: SANC at the one-loop level.}
\label{sanc-ann}
\end{center}
\end{figure}

Note, that all the numerical results of this section are produced with the so-called Standard SANC 
INPUT (section 6.2.3 of Ref.~\cite{Bardin:2005dp}).

%-----------------------------------------------------------------------------------------------------

\subsection{Decay channel $H \to f\bar{f} \gamma$ \label{decaychannel}}

For the decay channel we did not find in the literature complete one-loop calculations. 
We present here numerical results for the $H\to \mu^+\mu^-\gamma$ decay channel for $\mh = 150$ GeV. 

\vspace*{2mm}

$\bullet$ {\underline{GRACE, CompHEP and SANC at the Born level}\label{comp_born}}
 
\vspace*{2mm}
The results of the comparison for the total width in the Born approximation 
in GeV between GRACE Ref.~\cite{Yuasa:1999rg} CompHEP Ref.~\cite{Boos:2004kh} and SANC 
are shown in the Table \ref{BORNGCS}. Here the input parameters are as in CompHEP.

\begin{table}[!h]
\begin{center}
\begin{tabular}{|l|l|l|c|l|}
\hline
$E_\gamma$, GeV & $\Gamma$, GeV, \cite{Yuasa:1999rg}&$\Gamma$, GeV~\cite{Boos:2004kh}&$\Gamma$, GeV, SANC & \\
\hline
70     & 2.0490(1) & 2.0489(1) & 2.0491(1) &$\cdot  10^{-9}$\\
50     & 1.4187(1) & 1.4189(1) & 1.4188(1) &$\cdot  10^{-8}$\\
10     & 1.0029(1) & 1.0030(1) & 1.0030(1) &$\cdot  10^{-7}$\\ 
1      & 2.6265(2) & 2.6266(1) & 2.6264(1) &$\cdot  10^{-7}$\\ 
0.1    & 4.3329(2) & 4.3325(1) & 4.3326(1) &$\cdot  10^{-7}$\\
0.01   & unstable & unstable  & 6.0474(1) &$\cdot  10^{-7}$\\
\hline
\end{tabular}
\caption{Comparison for the total width 
between  Refs.~\cite{Yuasa:1999rg},~\cite{Boos:2004kh} and SANC in the Born approximation.}
\label{BORNGCS}
\end{center}
\end{table}

Note, that SANC produces stable results up to very small photon energies which is mandatory to have
correct description of soft and hard radiations.
\clearpage

$\bullet$ { \underline{SANC at the Born and one-loop levels} }
\vspace*{2mm}

\begin{figure}[!h]
\begin{center}
\includegraphics[height=12cm,width=7cm,angle=-90]{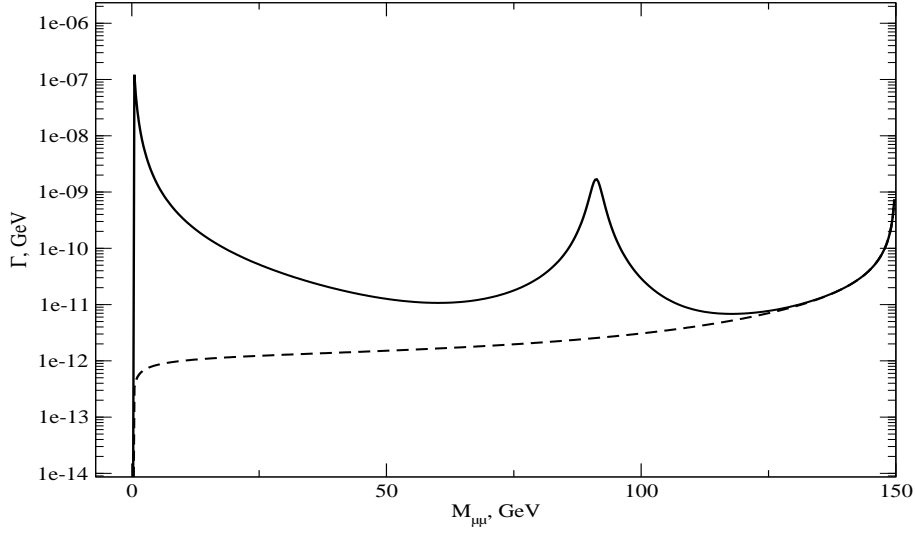}
\vspace*{-1mm}
\caption[Invariant mass distribution, Born (dashed) and one-loop levels (solid line)]
        {Invariant mass distribution, Born (dashed) and one-loop levels (solid line).}
\label{sanc-decay}
\end{center}
\end{figure}

In Fig.~\ref{sanc-decay} the fermion-antifermion invariant mass distribution
$d\Gamma/dM_{\mu^+\mu^-}$ is demonstrated.

\vspace*{9mm}

\vspace*{-2mm}

Two peaks due to $\gamma$ and $Z$ exchanges, are distinctly seen. The Coulomb peak region usually
does not represent any interest and should be cut out.

In Table~\ref{born-1-loop}, the partial width is shown in dependence on two cuts values, 
$s_{q\rm min,max},\;s_q=M_{\mu^{+}\mu^{-}}$ in the Born and one-loop approximation in two
schemes $\alpha$ and $G_{\mu}$.

\begin{table}[!h]
\begin{center}
\begin{tabular}{|c|r|r|c|c|}
\hline
 N  &$s_{q\rm min}$&$s_{q\rm max}$&$\Gamma^{\rm{Born}}\cdot 10^{-8}$
                                  &$\Gamma^{\rm{Born+1-loop}}\cdot 10^{-6}$\\
\hline
 &&&\multicolumn{2}{|c|}{$\alpha$ scheme}   \\
\hline
 1  &$2m_{\mu}$& 148.997  & 23.499 & 1.7536 \\
 2  &$2m_{\mu}$& 139.642  & 8.9737 & 1.6239 \\
 3  &$\mz -40$ &$\mz +40$ & 5.5040 & 1.2443 \\
 4  &$\mz -20$ &$\mz +20$ & 2.0188 & 1.1822 \\
\hline
 &&&\multicolumn{2}{|c|}{$G_{\mu}$ scheme}  \\
\hline
 1  &$2m_{\mu}$& 148.997  & 24.325 & 1.8152 \\
 2  &$2m_{\mu}$& 139.642  & 9.2891 & 1.6810 \\
 3  &$\mz -40$ &$\mz +40$ & 5.6975 & 1.2880 \\
 4  &$\mz -20$ &$\mz +20$ & 2.0898 & 1.2238 \\
\hline
\end{tabular}
\caption[SANC at the Born and one-loop levels]
        {SANC at the Born and one-loop levels.}
\label{born-1-loop}
\end{center}
\end{table}
\vspace*{-2mm}

All parameters and numbers are in GeV.
Two first $s_{q\rm max}$ are calculated in terms of $E_{\gamma}$ cut by the equation
$s^2_{q\rm max}=\mh^2-2\mh E_{\gamma}$
for $E_{\gamma}$=1 and 10 GeV, correspondingly.
As seen, the major part of the one-loop decay width is due to $Z$ resonance.
\clearpage

$\bullet$ {\underline{SANC in the resonance approximation at one-loop level} }
\vspace*{2mm}

The latter observation justifies to an extent the usual approach to the calculation of this decay,
the one-loop resonance approximation, which realized for example in PITHYA Ref.~\cite{Sjostrand:2006za}:
\bqa
\Gamma^{\rm{Res~1-loop}}_{H\to\mu^+\mu^-\gamma}=\frac{\Gamma^{\rm{1-loop}}_{H\to Z\gamma}
{\Gamma^{\rm{1-loop}}_{Z\to\mu^+\mu^-}}}{\Gamma_{\sss Z}}.
\eqa
In Table~\ref{resonance}, the total width is shown in dependence on cut value, 
$s_{q\rm min}$ in the resonance one-loop and in the complete one-loop approximations, 
again in two schemes $\alpha$ and $G_{\mu}$, here, however, without Born amplitude.

\vspace*{1mm}

\begin{table}[!h]
\begin{center}
\begin{tabular}{|c|c|c|c|c|}
\hline
$s_{q{\rm min}}$, GeV
&\multicolumn{2}{|c|}{$\Gamma^{\rm{Res~1-loop}}\cdot 10^{-6}$ GeV }
&\multicolumn{2}{|c|}{$\Gamma^{\rm{1-loop}}    \cdot 10^{-6}$ GeV } \\
\hline
         & $\alpha$&$G_{\mu}$& $\alpha$& $G_{\mu}$\\
\hline
$2m_\mu$ & 1.17006 & 1.2112  & 1.54394 &  1.59822 \\
1        & 1.17006 & 1.2112  & 1.45652 &  1.50773 \\
10       & 1.17006 & 1.2112  & 1.29776 &  1.34339 \\
30       & 1.16981 & 1.2109  & 1.22548 &  1.26857 \\
50       & 1.16771 & 1.2088  & 1.19604 &  1.23809 \\
70       & 1.15659 & 1.1973  & 1.17259 &  1.21381 \\
\hline
\end{tabular}
\caption[SANC, Resonance one-loop and complete one-loop approximations]
        {SANC, Resonance one-loop and complete one-loop approximations.}
\label{resonance}
\end{center}
\end{table}
Comparing columns computed in the same schemes, we see that the resonance approximation works 
with percent accuracy for strong cuts (50,70) GeV.

%--------------------------------------------------------------------------------------------------
\subsection{Channel $e\gamma\to e H$\label{comp_res_gammae_He}}

There is also reach literature devoted to this process (see, for example 
Refs.~\cite{Gabrielli:1997ix}--\cite{Banin:1998ap} and references therein).

We attempted a semi-tuned comparison of the total cross sections between
Table I of Ref.~\cite{Gabrielli:1997ix}
and SANC for three cms energies $\sqrt{s}=500,\;1000,\;1500\;$GeV and wide 
range of Higgs mass: 110~GeV$\leq\mh\leq$ 400~GeV. We tried to use all their masses
we manage to find in the paper and convention of coupling of the ``almost on-shell photon''. 

\begin{table}[!h]
\begin{center}
\begin{tabular}{|c|c|c|c|c|c|c|c|c|c|}
\hline
{$\mh/\sqrt{s}$} 
&\multicolumn{3}{|c|}{500}
&\multicolumn{3}{|c|}{1000}
&\multicolumn{3}{|c|}{1500}\\
\hline
&SANC&\cite{Gabrielli:1997ix}&$\delta$&SANC&\cite{Gabrielli:1997ix}&$\delta$&SANC&\cite{Gabrielli:1997ix}& $\delta$    \\
\hline
%--------------------------------------------------------------------------------
 80    &   8.40  &  8.38 &  -0.2 &   9.31 &   9.29 & -0.2 &    9.76 &  9.74 & -0.2   \\
 100   &   8.85  &  8.85 &  0    &   9.95 &   9.94 & -0.1 &   10.48 & 10.5  & -0.2   \\
 120   &   9.77  &  9.80 &  0.3  &  11.16 &  11.2  &  0.4 &   11.80 & 11.8  &  0     \\
 140   &  11.76  & 11.8  &  0.3  &  13.68 &  13.7  &  0.1 &   14.52 & 14.6  &  0.6   \\ 
 160   &  20.91  & 21.1  &  0.9  &  24.82 &  25.0  &  0.7 &   26.48 & 26.6  &  0.5   \\
 180   &  20.67  & 20.9  &  1.1  &  25.04 &  25.3  &  1.0 &   26.81 & 27.0  &  0.7   \\
 200   &  16.99  & 17.2  &  1.2  &  21.05 &  21.2  &  0.7 &   22.64 & 22.8  &  0.7   \\
 300   &   5.90  &  5.97 &  1.2  &   8.44 &   8.53 &  1.0 &    9.33 &  9.43 &  1.1   \\
 400   &   1.64  &  1.64 &  0    &   2.74 &   2.78 &  1.5 &    3.15 &  3.18 &  1.0   \\
\hline
\end{tabular}
\caption[$H$ production channel, SANC and  Ref.~\cite{Gabrielli:1997ix}]
        {$H$ production channel, SANC and  Ref.~\cite{Gabrielli:1997ix}.}
\end{center}
\end{table}

In the Table we show total cross sections $\sigma$ and relative difference $\delta$ 
between two calculations ($\delta={\sigma\cite{Gabrielli:1997ix}}/{\sigma[\rm{SANC}]}-1,\,(\%)$).
As seen, the difference in the vast majority of points is below 1\% and shows
up irregular behavior pointing to its numerical origin
(our numbers are calculated with real*16).
Given these observations, we consider the two results to be in a very good agreement.

%----------------------------------------------------------------------------
\def\href#1#2{#2}
\end{document}